\documentclass[12pt]{article} 

\usepackage{graphicx}
\usepackage{txfonts}
\pdfoutput=1
\begin{document}

\begin{center} 

{\Large \bf  Diffusion in barred-spiral galaxies}

\bigskip
Maura Brunetti$^1$, Cristina Chiappini$^{1,2}$, and Daniel Pfenniger$^1$\\

\bigskip
\noindent {\small
$^1$Geneva Observatory, University of Geneva, CH-1290 Sauverny, 
Switzerland}\\ 

\noindent 
{\small $^2$INAF, Osservatorio Astronomico di Trieste, Via G. B. Tiepolo 
11, 34100 Trieste, Italy}

\end{center}

\bigskip
\abstract{We characterize the radial migration of stars in the disk plane 
  by calculating the diffusion coefficient and the diffusion
  time-scale for a bulge-disk $N$-body self-consistent system 
  with a marginally-stable Toomre-Q parameter. 
  We find  that diffusion is not constant in time, but follows the evolution of the bar, and 
  becomes maximum near the corotation region and in the external disk region, where asymmetric patterns develop. 

  }


\section{Introduction}\label{introduction}

The majority of
local galaxies are barred-spiral galaxies, such as our Milky Way~\cite{delgado,eskridge} where  
the bar is the strongest non-axisymmetric pattern in the disk. The bar is
generally coupled to other non-axisymmetric patterns such as
spirals and warps if the disk is sufficiently cold~\cite{revaz}. 
The bar has a time-dependent activity, with a pattern speed which 
typically decreases in isolated galaxies.  
However, the system can be cooled again by adding dissipative 
infall of gas, or by forming stars on low-velocity dispersion orbits, with the net effect of  
restoring the amplitude of spiral waves and 
the strength of the bar, or even destroying it. In this way bars (and
spiral waves) can be seen as recurrent patterns which can be rebuilt
during their long history until the present configuration at $z=0$~\cite{bournaud}. 

Under the action of these non-axisymmetric patterns, stars move in
the disk which gradually becomes hotter.    
Velocity dispersion of disk stars rises with age, 
as confirmed by observations in the 
Solar neighborhood (see~\cite{binney00} and references therein) and in
external galaxies~\cite{gerssen, shapiro}. The origin and the amount of 
disk heating are still open to debate. 

First attempts to explain such a heating process in the disk of
galaxies tried to model empirically the observed increase of the
stellar velocity dispersion with age in the solar neighborhood. 
Wielen~\cite{wielen} suggested a diffusion mechanism in velocity space, which
gives rise to typical relaxation times for young disk stars of the order of
the period of revolution and to a deviation of stellar positions 
of 1.5~kpc in $200$~Myr. The result was obtained without making
detailed assumptions  on the underlying local acceleration process
responsible for the
diffusion of stellar orbits. Global acceleration processes, such as the 
gravitational field of stationary density waves or of central bars
with constant pattern speed,
were ruled out since their contribution 
to the velocity dispersion of old stars is negligible and concentrated
in particular resonance regions~\cite{wielen,binneytremaine}. 
Different local accelerating mechanisms have been investigated so far
in isolated galaxies, 
such as the gravitational encounters
between stars and giant molecular clouds~\cite{spitzer51, spitzer53,lacey}, 
secular heating produced by transient 
spiral arms~\cite{barbanis,carlberg,fuchs} or
the combination of the two processes~\cite{binney88, jenkins}.

The radial excursion predicted by Wielen~\cite{wielen} is not sufficient to
explain the weakness of the correlation between age and metallicity
in the Solar neighborhood (see, for example, \cite{edvardsson}).
In order to explain both the large scatter in the 
age-metallicity relationship and the evidence that even old disk
stars today have nearly circular orbits, Sellwood \& Binney~\cite{sellwood} have
recently suggested a new mechanism based on the resonant
scattering of stars under the effect of transient spiral waves. 
In this process,
a star initially on a nearly circular orbit resonates with a rotating
wave and changes its angular momentum. If the duration of the peak
amplitude of the perturbing potential is less than the period of the
`horseshoe' orbits, i.e.~orbits of particles trapped at the
corotation radius of the spiral wave, the star can escape from the
potential well without changing its eccentricity. 
The net effect of this scattering mechanism is that stars migrate
radially without heating the disk. In other words, the overall distribution of
angular momentum is preserved, except near the corotation region of
the transient spiral wave, where stars can have large changes of their
angular momenta. 

The mechanisms driving radial migration and heating are still 
hotly debated. Minchev et al. ~\cite{minchev10a, minchev10b} propose a mixing mechanism where resonances between the 
bar and the spiral arms can act much more efficiently than transient spiral structure, 
dramatically reducing the predicted mixing timescales.     

Most of the observational signatures of radial mixing reported in the literature~\cite{grenon72, castro, grenon99} point to stars coming from a region next to the bulge/bar intersection. Radial 
migration of the stars in the disk have been
suggested by Haywood~\cite{haywood},  who estimated upper values for the migration rate from 1.5 to 
3.7 kpc/Gyr, which agree with the values estimated in~\cite{lepine} for the radial wandering due 
to the scattering mechanism assumed by Sellwood and Binney~\cite{sellwood}.  High resolution 
cosmological simulations~\cite{roskar} confirm that such scattering mechanism determines a 
significant migration
in the stellar disk. However these simulations ignore the important effects of bars found 
by Minchev et al. ~\cite{minchev10a, minchev10b}. Radial migration of stars (and gas) could have important implications 
for the interpretation of key observational constraints, such as the age-metallicity 
relationship or the metallicity gradients, since old stars that formed at small 
galactocentric radii from enriched
gas or young metal-poor stars at large radii are enabled to appear in a 
Solar-neighborhood sample. Due to the lack of detailed information on the process driving 
radial mixing, models of the Galactic chemical evolution have evaluated past history of 
the solar neighborhood and the formation and evolution of the abundance gradients 
assuming that radial mixing did not played an important role~\cite{vandenbergh, schmidt, pagel, chiappini97, chiappini01}. 
Recently, Schoenrich and Binney~\cite{schonrich08} explored the consequences of mass exchanges between
annuli by taking into account the effect of resonant scattering
of stars described before. This approach appears to be successful to replicate many 
properties of the thick disk in the Solar neighborhood without requiring any merger or 
tidal event~\cite{schonrich09}. However, again in this case, the strong mixing mechanisms driven by bar resonances were not taken into account, casting thus doubts on some of their 
conclusions.

In order to include the effect of radial migration in 
chemical evolution models and to gain a global (chemical +
kinematical) vision of the
processes at play in the galactic disks, many dynamical 
aspects need to be further investigated and in particular 
the role of the bar, that is the main non-axisymmetric component 
in disk galaxies. This is what we start to do in the 
present paper. In particular, we consider a marginally-stable disk which develops a
central bar and spiral arms with the aim of quantitatively estimating 
the time and length scales of star 
diffusion in the radial direction. The final goal is to investigate how these
characteristic scales evolve in time and how they
depend on the activity of the bar. We present in detail the methods used for calculating the diffusion coefficient and time-scale. We will apply these methods to models which include the halo component and disks with different grade of stability in a forthcoming paper.     

The paper is organised as follows. In section~2 we describe the
simulation and the relevant parameters. In section~3 we solve the
diffusion equation in axisymmetric systems, we define the diffusion
coefficient, the diffusion time-scale and the diffusion length-scale,
and we describe how we
estimate these quantities from the simulation results. In section~4 we
present our results and in section~5 we summarise our findings.  

\section{$N$-body simulation}

We have run self-consistent $N$-body simulations starting from a 
bar-unstable axisymmetric model. The initial configuration is marginally
stable, i.e. the initial
value of the Toomre parameter $Q_T = \sigma_r\, \kappa/ (3.36\, G\,
\Sigma)$ is $Q_T \sim 1$, 
where $\sigma_r$ is the radial velocity dispersion of the
disk component, $G$ is the gravitational constant, 
$\Sigma$ is the disk surface density, 
$\kappa$ is the epicycle frequency defined by 
$\kappa^2 = R d\Omega^2/dR + 4\Omega^2$, where $\Omega$ is the circular
frequency related to the global gravitational potential $\Phi(R,z,t)$
in the disk plane $z=0$ by 
$\Omega^2 = (1/R)\, \partial \Phi/\partial R$. 

The initial mass distribution in our
simulations corresponds to a superposition of a pair of axisymmetric
Miyamoto-Nagai disks of mass $M_B$, $M_D$, horizontal scales $A_B+B$,
$A_D+B$, and identical scale-height $B$, 
\begin{equation}\label{MN}
\Phi_{MN}(R,z)=\sum_{i=B,D} \frac{-GM_i}{\sqrt{R^2+(A_i+\sqrt{B^2+z^2})^2}}
\end{equation}
The first component represents the bulge ($B$), while the
second the disk ($D$)~\cite{pfenniger91}. The parameters have been set to
$A_B = 0.07$~kpc, $A_D = 1.5$~kpc, $B=0.5$~kpc, $M_B/M_D = 3/17$. The
initial particle positions and velocities are found by a pseudo-random draw 
following the density law corresponding to eq.~(\ref{MN}),
truncated to a spheroid of semi-axes $R=30$~kpc, $z=10$~kpc. 
The number of particles in the disk-bulge component is $N = 4\cdot
10^6$. 
We then solve for the first velocity moment equations (Jeans equations) to find 
an approximate local equilibrium. The resulting distribution is then 
relaxed for a couple of rotations until ripples spreading through the 
disk from the center disappeared. 
We used this as the initial condition for the 
$N$-body simulations, performed with the {\tt Gadget-2} code~\cite{springel01, springel05}. 

\begin{figure*}
   \centering
   \includegraphics[width=5.2cm]{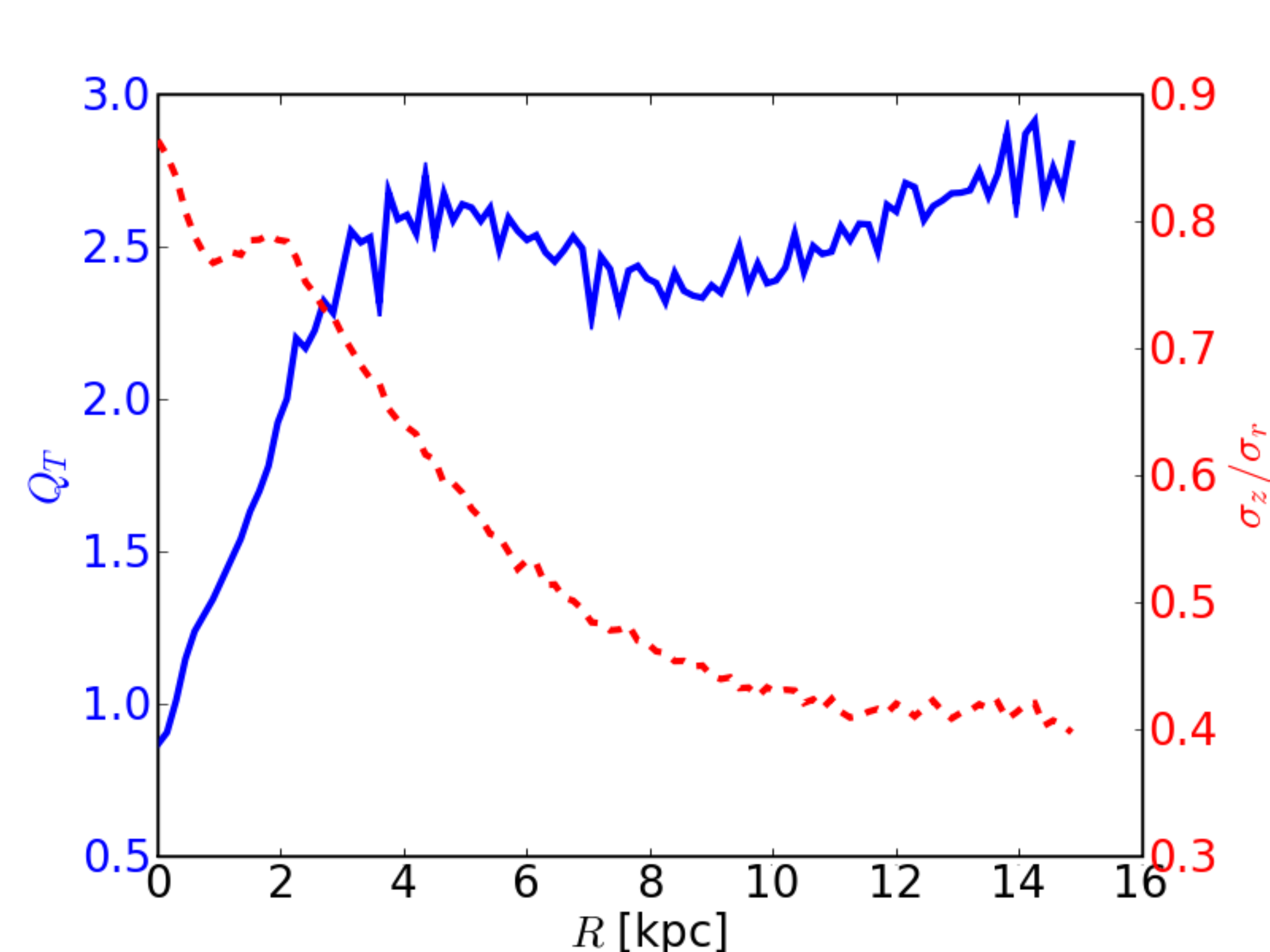} 
   \includegraphics[width=5.2cm]{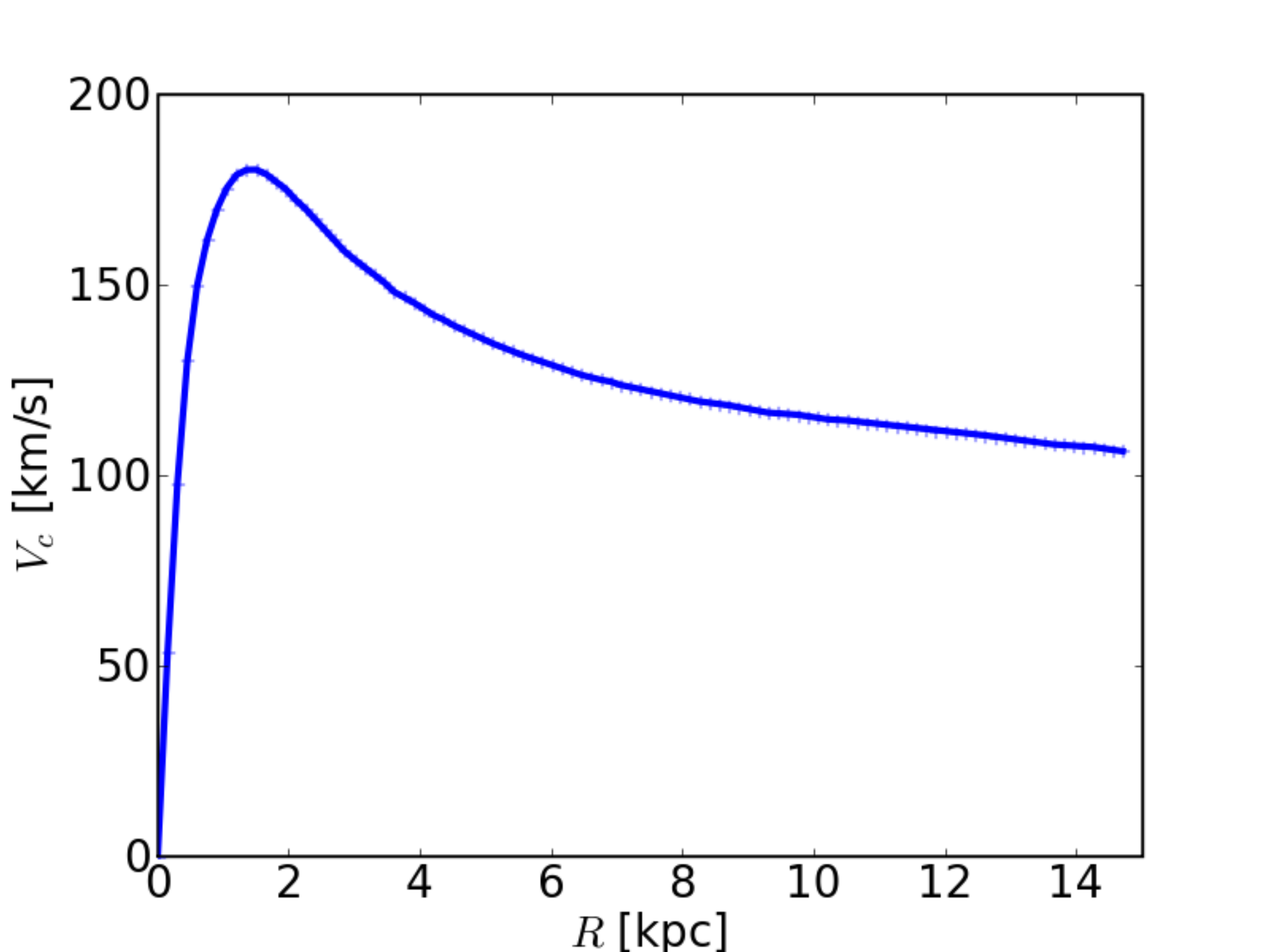} 
   \includegraphics[width=5.2cm]{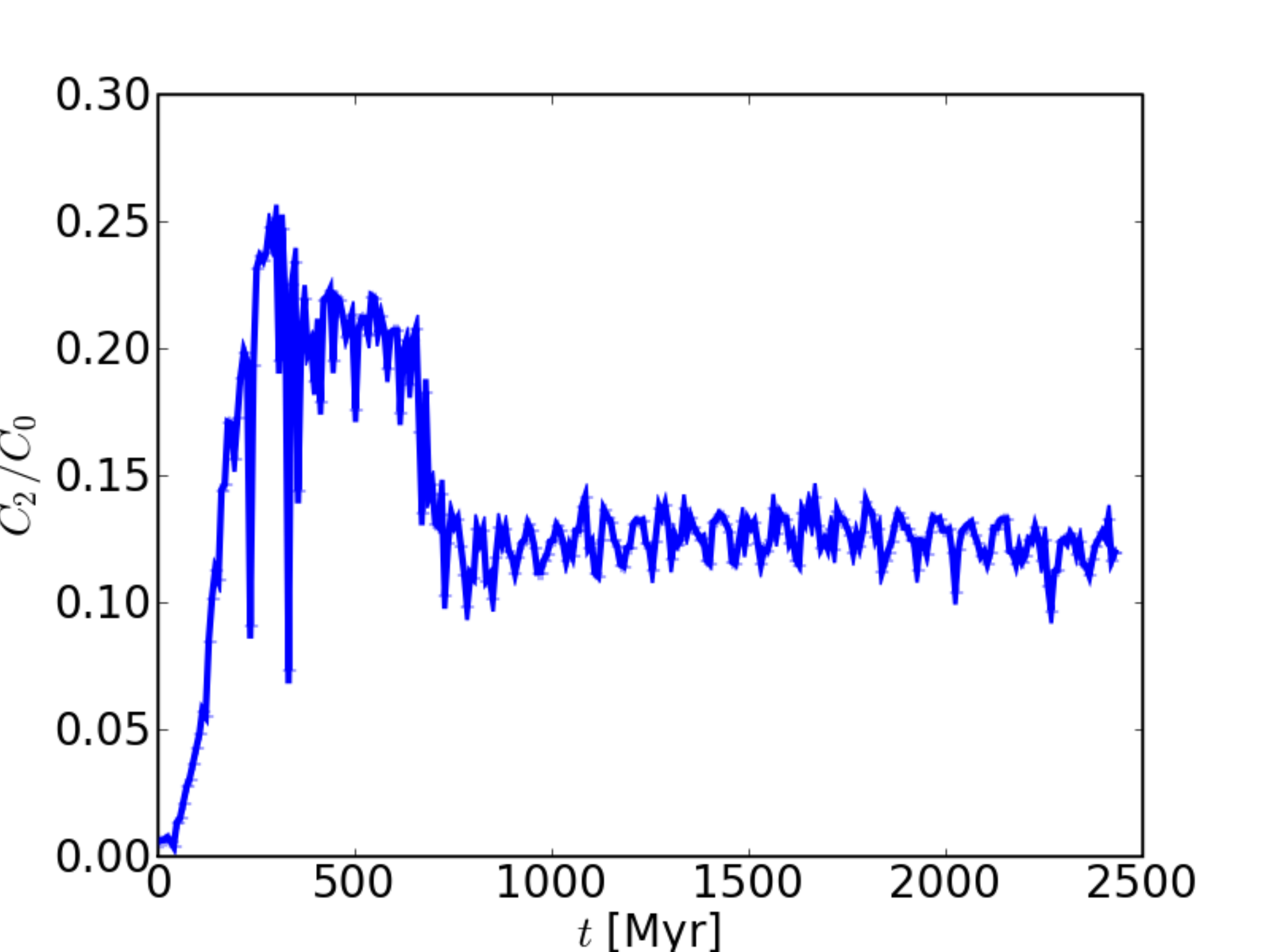} 
   \caption{\small {\it Left}:
   Toomre-parameter (solid blue line) and Araki ($\sigma_z/\sigma_r$) parameter (dashed red line, labels on the right $y$-axis) at the final time. {\it Middle}: rotation curve. {\it Right} bar-strength time evolution.}
   \label{RotCurve}%
\end{figure*}

The initial value of
the radial and vertical velocity dispersions at two scale lengths from
the center is $\sigma_r \sim \sigma_z \sim 20$~km/s. 
The disk is
marginally stable and spiral waves develop 
with the main global effect of
heating in the radial direction. 
The Toomre and Araki parameters at the final time
are shown in the left panel  of 
fig.~\ref{RotCurve}, solid blue and dashed red lines, respectively. 
The rotation curve at the final time is shown  
in the middle panel of fig.~\ref{RotCurve}.

We classify stellar orbits into three dynamical categories~\cite{sparke,pfenniger91}: 1) the bar orbits
and the disk orbits with the Jacobi integral $H=E-\Omega_p L_z$ smaller
than the value at the Lagrangian points $L_{1,2}$, $H<
H(L_{1,2})$, where $E$ is the total energy and $L_z$ is the $z$-component of
the angular momentum. The particle separation in the bar or disk
component can easily be done since the bar orbits have typically
smaller values of $L_z$ and $E$. 2) The hot orbits with $H\ge H(L_{1,2})$. 

The bar pattern speed $\Omega_P \equiv d\theta/dt$, where 
$\theta$ is the azimuthal angle of the bar major axis (in the inertial 
frame) calculated by diagonalising the moment of inertia tensor of the 
bar particles, is $40~\rm{km}~\rm{s}^{-1}~\rm{kpc}^{-1}$.
We find that the pattern speed slowly decreases in time with a 
rate of a few km/s/kpc/Gyr in agreement with~\cite{fux,bournaud}. 
The corresponding corotation 
radius $R_c$, obtained by the intersection of $\Omega_P$ with the 
circular frequency, $\Omega_P(t) = \Omega(R_c,t)$, increases in time 
(it ranges from $R_c = 2$ to 4\, kpc in our simulation). 

In order to understand the role played by the central bar on the
distribution of stars in the disk, we follow the
evolution of the bar strength in time. The amplitude $C_m $ of the mode $m$ in  the density distribution is defined as
\begin{equation}
C_{m} = \left |\sum_{j} \exp(i\, m\theta_j)\right |,\quad
\end{equation}
where $j$ labels the stars.
The bar's strength  is defined as the mode $C_{m=2}$ when the   stars are restricted to  the bar component. 
We normalise $C_2$ with respect to the number of stars in the
bar component, $C_0$. This quantity is shown in the right panel of fig.~\ref{RotCurve}.
Since we do not include the gas component in our model, we are not
able to follow its evolution for times longer than few Gyr. After
this typical time-scale, the bar amplitude saturates and the system reaches
a quasi-steady state. Since the inclusion of the gas component necessarily 
requires the introduction of other less controlled parameters, 
such as the cooling rate or star formation, we prefer to limit the integration 
time and to study the physical process at play within few Gyr.

\begin{figure*}
  \centering
   \includegraphics[width=7cm]{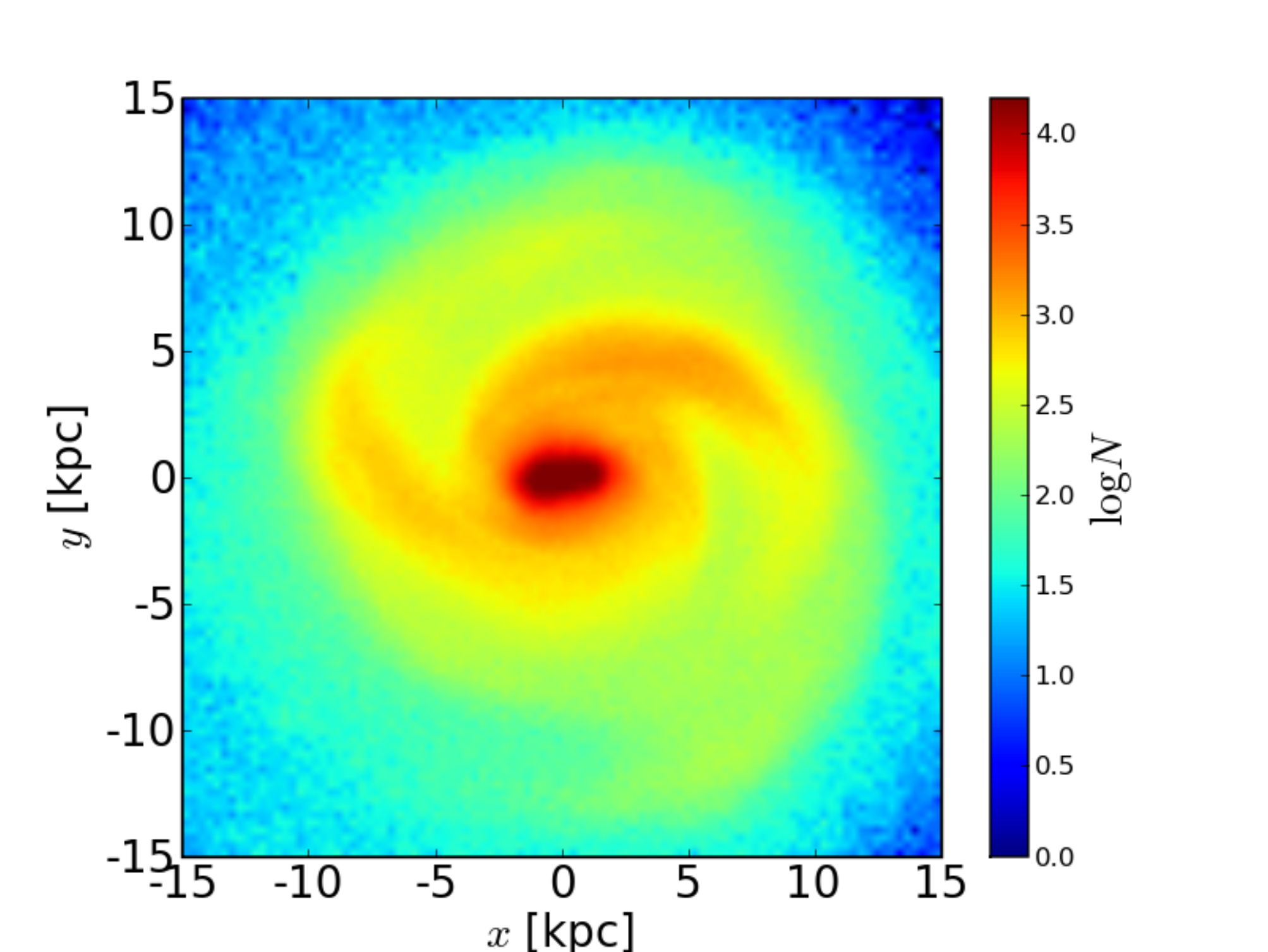} 
   \includegraphics[width=7cm]{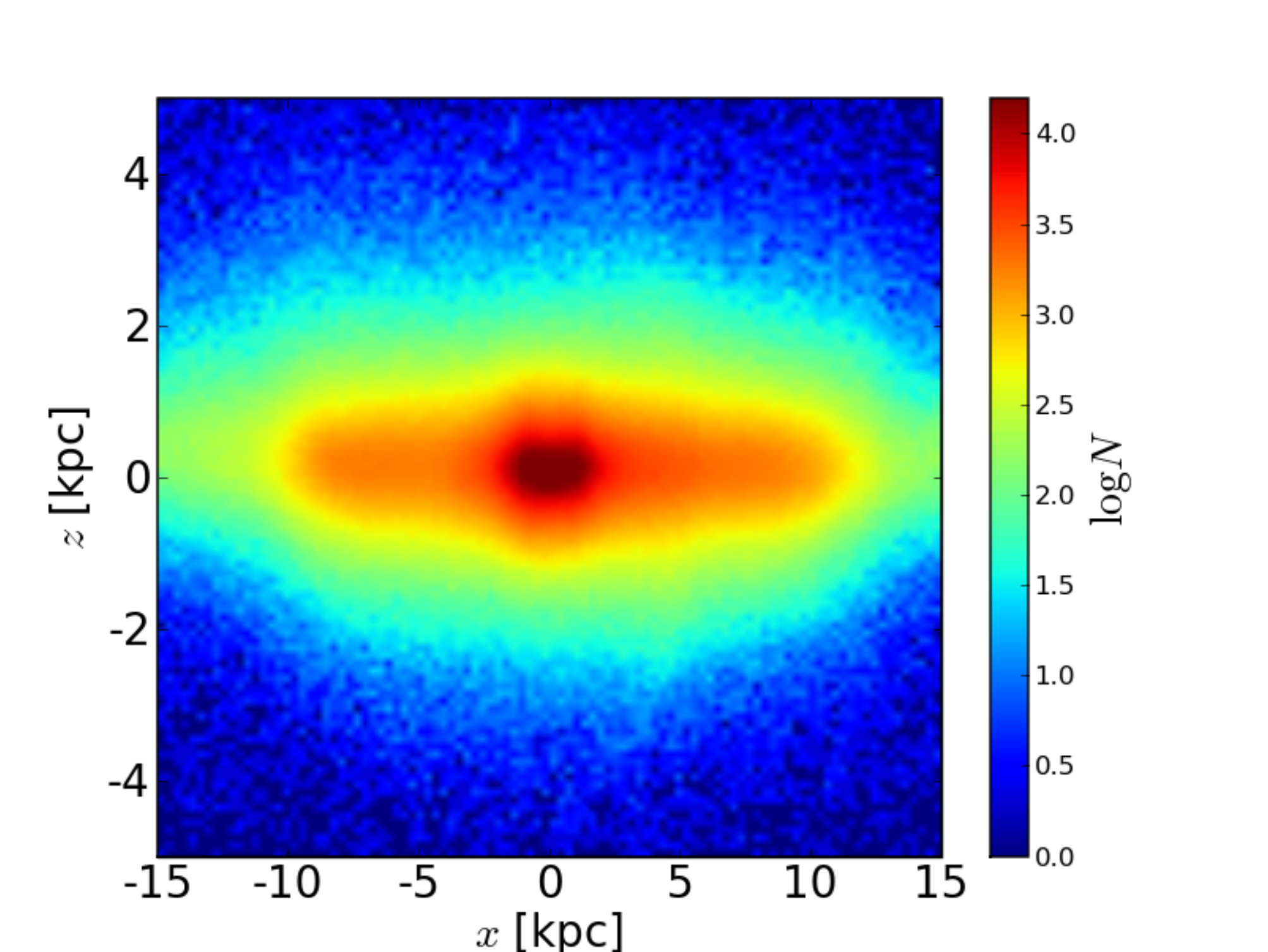} 
   \caption{Density maps at $t\sim 550$~Myr. {\it Left}: face-on
   view. {\it Right}: edge-on view.}
   \label{figDensity}
\end{figure*}

The face-on and edge-on views of the density distribution 
are shown in 
fig.~\ref{figDensity} at time $t \sim 600$~Myr, that is just after the moment when
the bar's strength reaches its maximum. Both bar and 
spiral arms develop, since
the disk is sufficiently cold. 
In the external regions, the dominant pattern has $m =1$.

\section{Diffusion equation in axisymmetric systems}

\begin{figure}
   \centering
   \includegraphics[width=7cm]{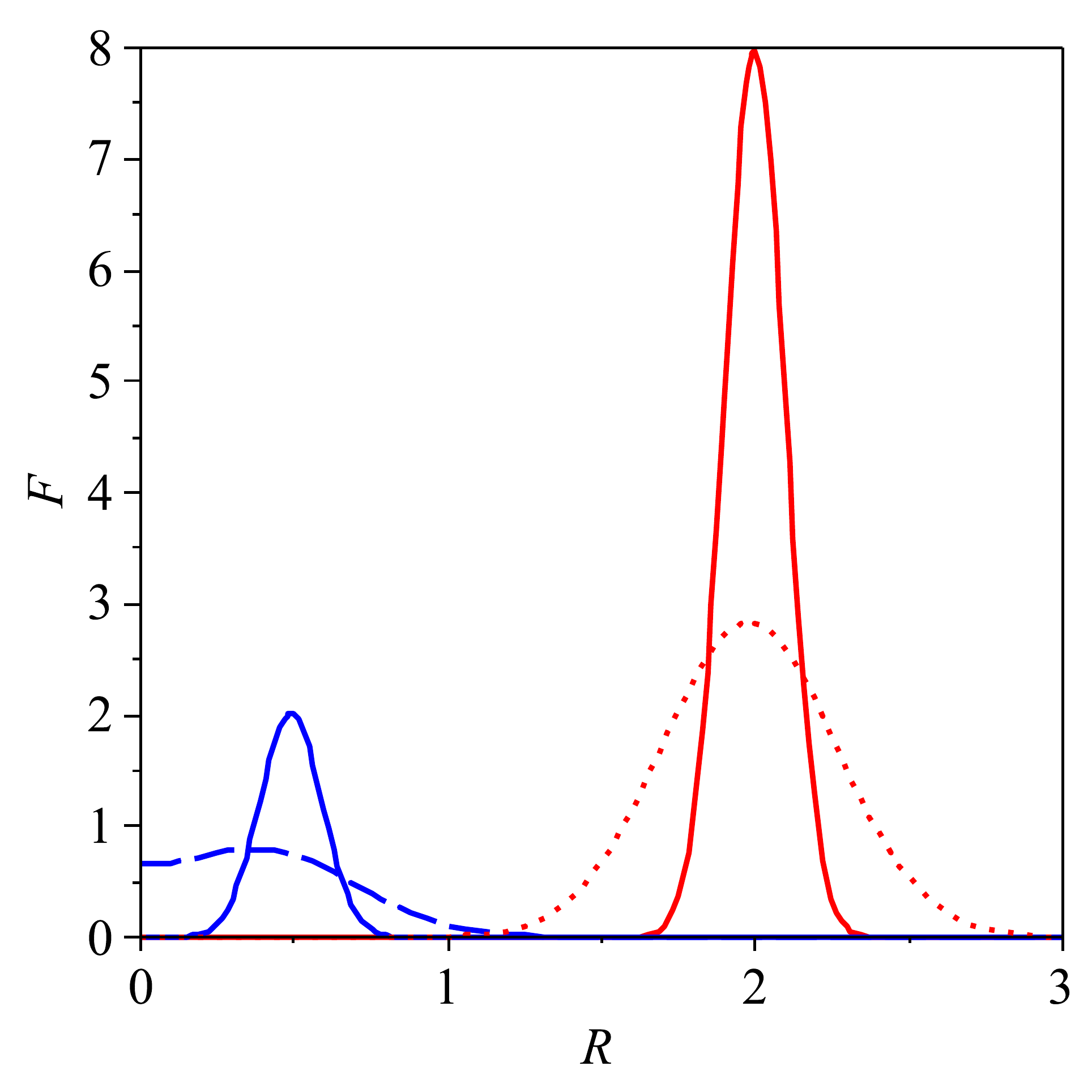} 
     \caption{\small Two distributions with $D = 1$ and centered in
   $R_0 = 0.5$ and 2 (solid lines) evolve in time (dashed and dotted
   lines, respectively), as described by
   eq.~(\ref{eqDiff}). }
    \label{fig:solution}
\end{figure}

We model the flow along the radial direction in the galactic plane 
by introducing the function
$F(R,t)$ which satisfies the {\it spatial} diffusion equation 
\begin{equation}
\partial_t F = \frac{D}{R}\, \partial_R (R\partial_R F)\, ,
\label{DiffEq}
\end{equation}
where $D$ is the diffusion coefficient.
The general solution of
eq.~(\ref{DiffEq}) which is non-singular at $R=0$ is given by 
\begin{equation}
F(R,t) = \int_0^\infty A(s)\, e^{-D t s^2} J_0(sR)\,s\, ds\, ,
\label{sol1}
\end{equation}
where $J_0(x) = J_0(-x)$ is the Bessel function of the first kind. 
The function $A$ can be
determined by taking the Hankel transform of $F(R,0)$
\begin{equation}\label{Hankel}
A(s) = \int_0^\infty F(R,0)\, J_0(sR)\, R\, dR\, .
\end{equation}
Thus, by inserting eq.~(\ref{Hankel}) into eq.~(\ref{sol1})  and assuming that the particles are initially localized at a certain radius $R_0$ at time $t_0 =0$, $F(R,0) = F_0R_0\delta(R-R_0)$, we obtain
\begin{equation}
F(R,t) = R_0^2 F_0 \int_0^\infty s\, e^{-D t s^2} J_0(sR)\, J_0(sR_0)\,
ds\, .
\end{equation}
Thus, the diffusion of this distribution can be expressed in term of elementary and Bessel 
functions~\cite{gradsteyn}. The time-evolution of an initial set of localized particles reads 
\begin{equation}\label{eqDiff}
F(R,t) = \frac{R_0^2 F_0}{2D t} 
\exp\left(-\frac{R_0^2+R^2}{4D t}\right)\,
I_0\left(\frac{RR_0}{2D t}\right)\, , 
\end{equation}
where $I_0(x)$ is the modified Bessel function of the first kind, which
is finite at the origin, $I_0(0)=1$.  In fig.~\ref{fig:solution} 
two initial distributions with $D = 1$ and centered in $R_0 =
0.5$ and 2 (solid lines) evolve in time, as described by
eq.~(\ref{eqDiff}) (dashed and dotted lines, respectively). At large
radii, the distributions are essentially Gaussian, while at small
radii they are strongly modified from the contributions of particles
at the center of the cylinder.     
Eq.~(\ref{eqDiff}) describes 
the distribution of the radial positions of stars in the disk  
at the initial time $t_i$ which diffuse toward position
$R_0$ at time $t_0$, such that $t_0 - t_i = \Delta t \le T_D$, where $T_D$ 
is the diffusion time-scale or, equivalently, the distribution of stars 
which initially are in $R_0$ at $t_0$
and then diffuse toward $R$ with $\Delta t \le T_D$.

For $R$ which goes to zero, eq.~(\ref{eqDiff}) reduces to 
\begin{equation}
F(0,t) = \frac{R_0^2 F_0}{2D t} \, \exp\left[-R_0^2/(4D
  t)\right]\, .
\end{equation} 
For large values of the argument the modified Bessel function 
$I_0(x) \to (2\pi x)^{-1/2}\,  \exp(x)$ and thus $F(R,t)$ reduces to 
\begin{equation}\label{limEqDiff}
\lim_{R\to \infty} F(R,t) = \frac{R_0^{3/2} F_0}{\sqrt{4\pi D t R}} 
\exp\left(-\frac{(R-R_0)^2}{4D t}\right) = \frac{R_0^{3/2}
  F_0}{\sqrt{R}}\, {\cal{N}}(\mu,\sigma) 
\end{equation}
where ${\cal N}(\mu, \sigma)$ is the Gaussian distribution with mean
value $\mu = \langle R\rangle  = R_0$ and 
standard deviation $\sigma = \sqrt{2D t}$. 

As we will describe in the following
subsection, the $N$-body simulation presented in 
section~2 allows us to estimate the diffusion coefficient
$D$ and the diffusion time-scale $T_D = |t_0-t_i|$ in marginally stable 
disks. 

\subsection{Calculation of diffusion time-scale and diffusion coefficient}

The diffusion coefficient $D$ in
eqs.~(\ref{DiffEq})-(\ref{eqDiff}) can be regarded as an instantaneous 
coefficient which depends on the position at which particles are 
initially localised and on the diffusion time. Modeling of 
stellar migration as a diffusion process is valid only for time 
intervals less than the diffusion time-scale, $\Delta t \le T_D$, and
for time intervals larger than the typical chaotic time-scale.     
Thus, the first issue is to estimate this diffusion time-scale from
the simulation results. In the right panel of fig.~\ref{fig:solution}, we show the radial 
dispersion $\langle \Delta R^2 \rangle$ of stars initially at $(R_0,
t_0)$ as a function of time, for different values of $(R_0,t_0)$. 
At $t=t_0$ the radial dispersion is null
since the particles are all localised at $R=R_0$. 
As time evolves,  $\langle \Delta R^2 \rangle$
grows with a linear rate and then it saturates after a diffusion
time-scale $T_D$.  The diffusion time scale turns out to be 
of the order of the rotation period, $T_D \sim T_{rot} = 2\pi/
\Omega(R,t)$. 

\begin{figure}
   \centering
   \includegraphics[width=7cm]{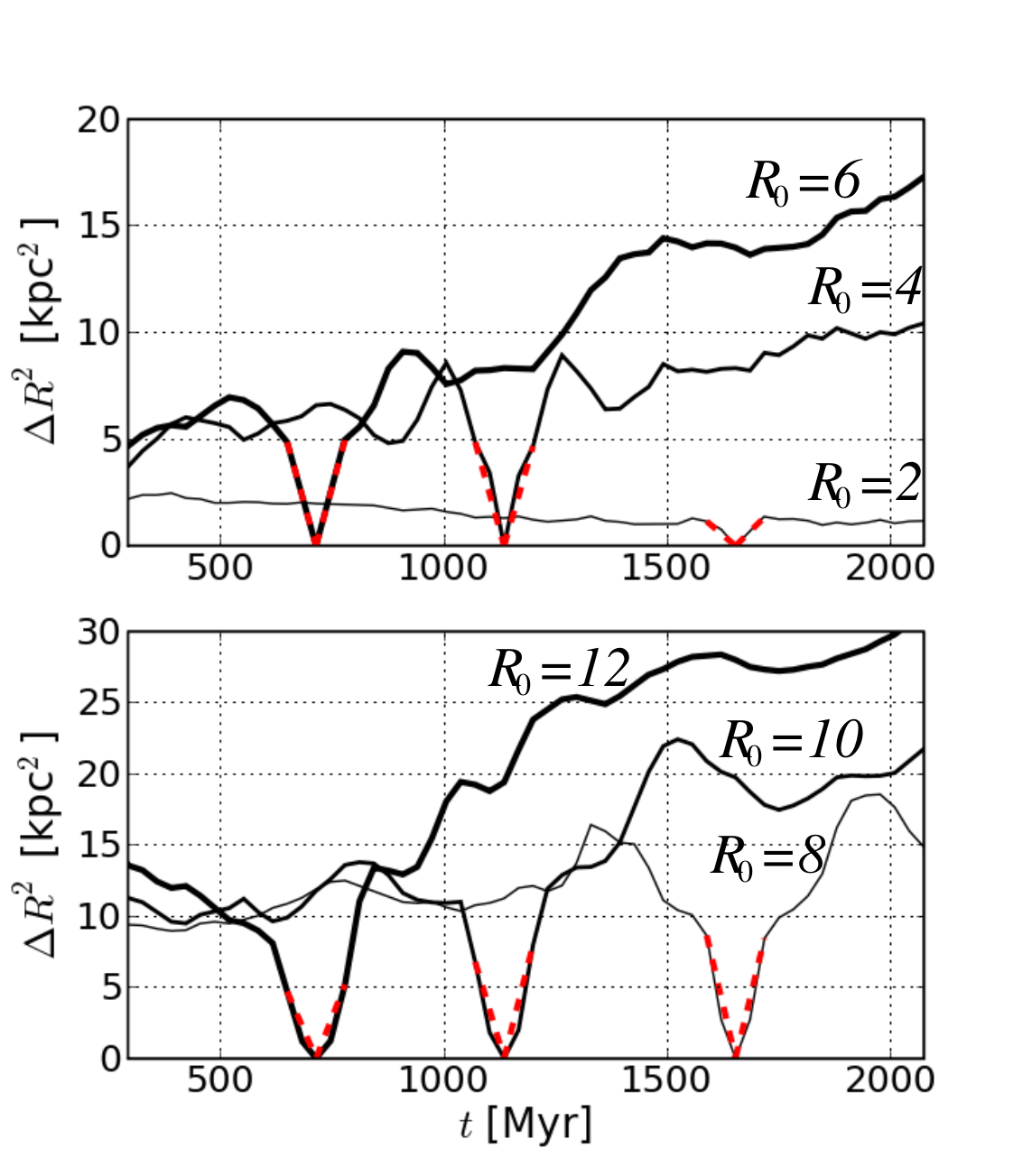}  
   \includegraphics[width=8cm]{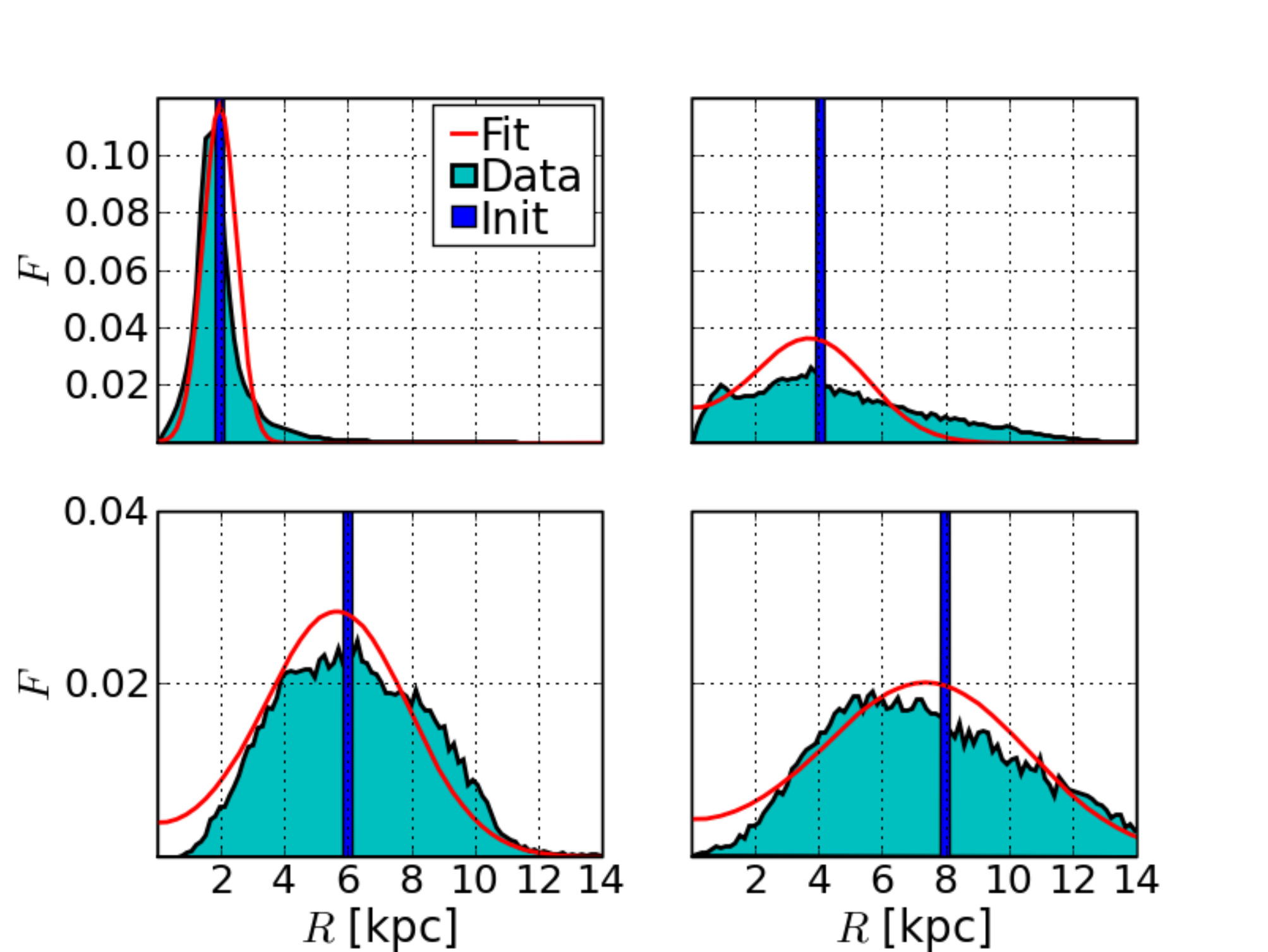} 
   \caption{\small {\it Left:} radial dispersion of particles initially at
   $(R_0,t_0)$, for $R_0 = 2,4,\ldots, 12$~kpc and $t_0 = 700, 1150,
   1650$~Myr, as a function of time. The growth rates in the vicinity
   of the initial time $t_0$ give a first estimate of the diffusion
   coefficient from the relation $\langle \Delta R \rangle^2 =
   \sigma^2 = 2D t$ (dashed red lines).  {\it Right}: distribution of particles at radius $R_0=2,4,6,8$~kpc and different initial times $t_0$ (dark blue), distribution of the same particles at time t, such that $t-t_0 < T_D$ (light blue) and fit calculated by the nonlinear lest-square method (red line).} 
   \label{fig:fit}
\end{figure}

From the relation $\langle \Delta R^2 \rangle = \sigma^2 = 2D t$,
we calculate a first estimate of the diffusion coefficient $D$ 
as the growth rates in the vicinity of the initial time $t_0$ 
(red dashed lines in the left panel of fig.~\ref{fig:solution}) 
and we use these values as initial guess of the nonlinear least-square
method, which minimizes the difference between the 
numerical results and the general solution of the
diffusion equation described by eq.~(\ref{eqDiff}) for times less than
$T_D$. The maximum relative error on the estimated parameter $D$ is $\Delta D/D \sim 0.06$
near the central region.  The fitting function that we obtain by this procedure 
reproduces reasonably well the overall distribution of particles at times 
$t>t_0$ (or $t<t_0$), with $\Delta t = |t-t_0|< T_D$, as can be seen in the right panel of fig.~\ref{fig:fit}.
  
\section{Results}

\begin{figure*}
   \centering
   \includegraphics[width=7cm]{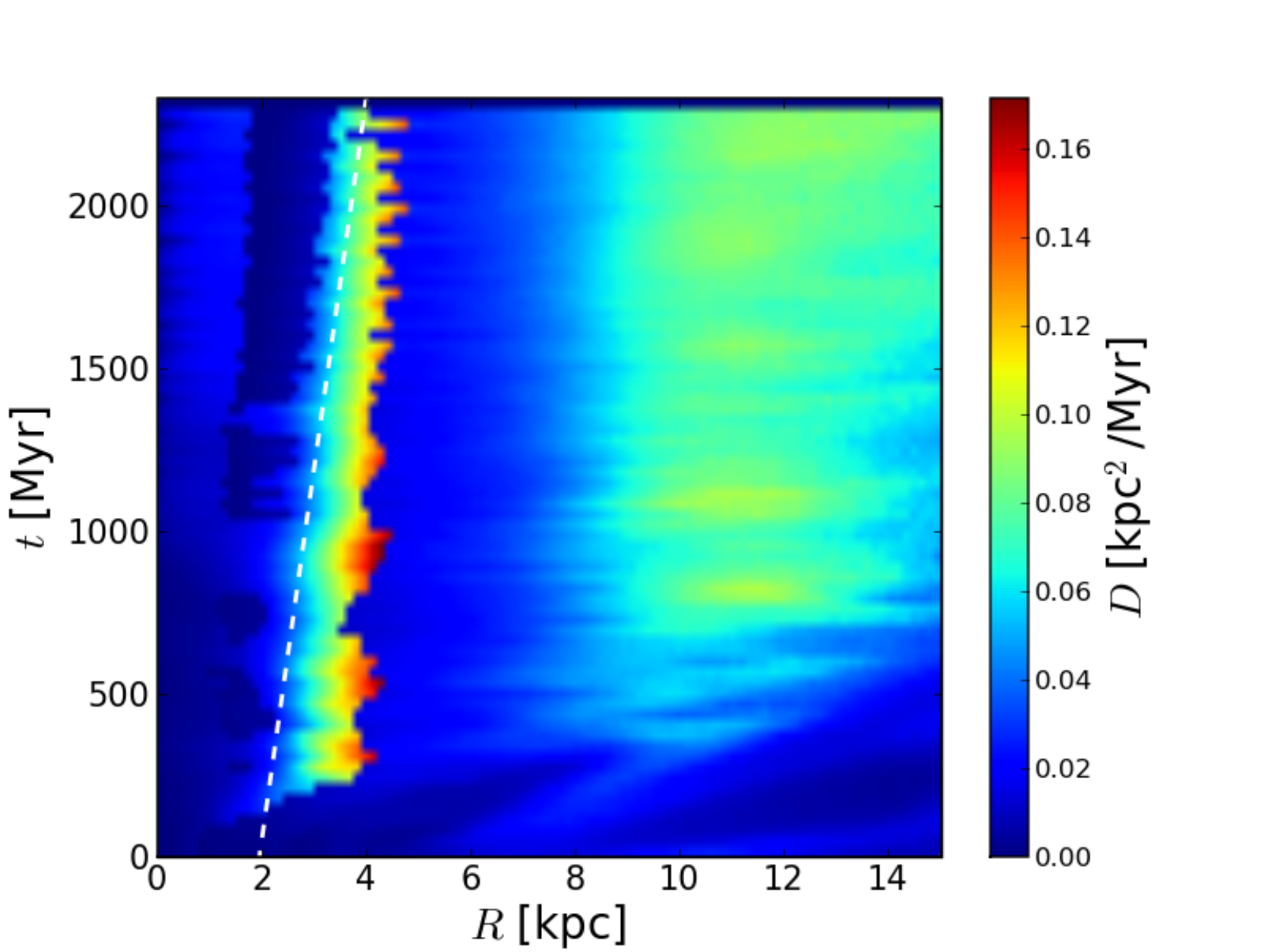} 
   \includegraphics[width=7cm]{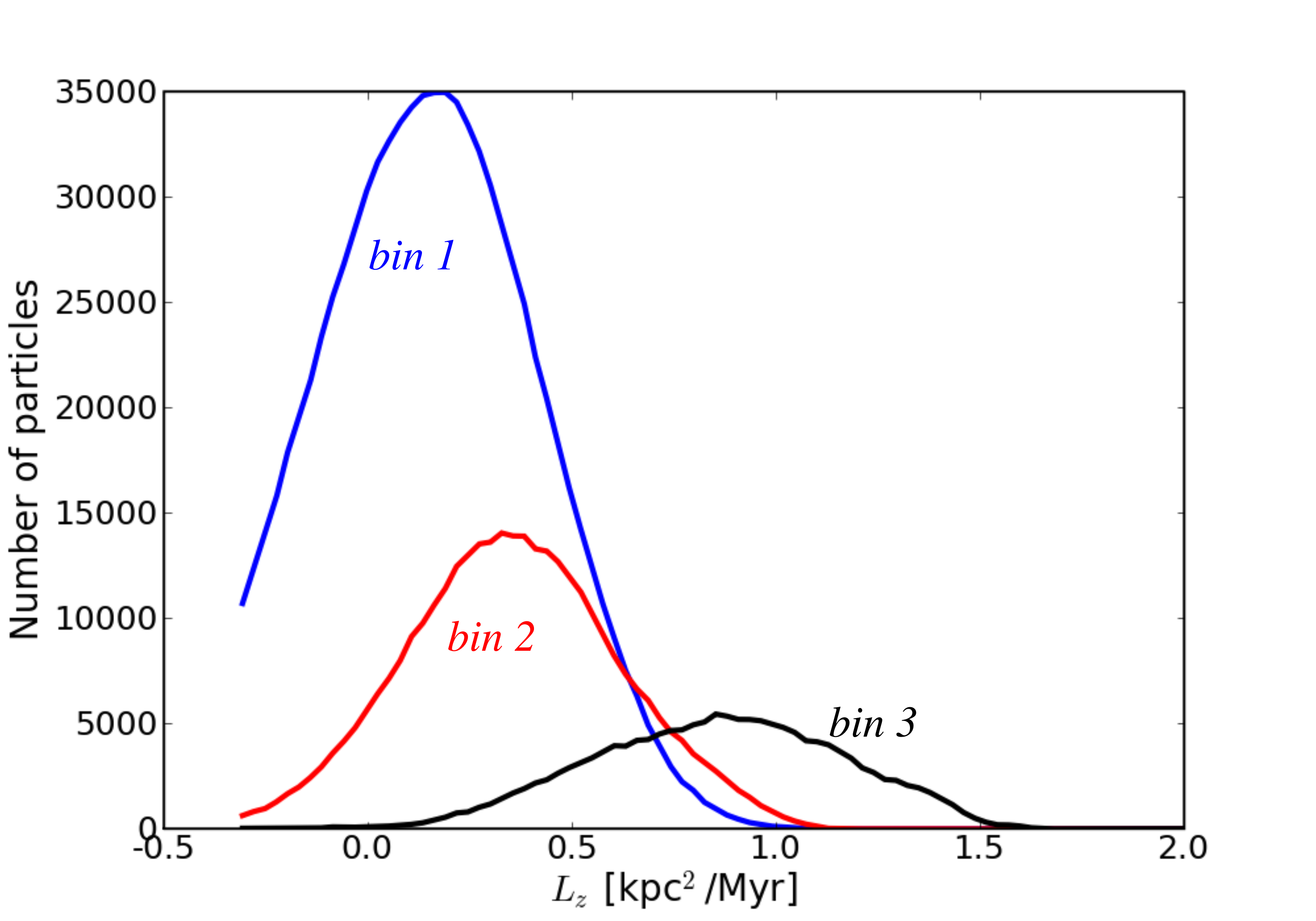} 
   \caption{\small {\it Left:} contour map of the diffusion
     coefficient $D$. {\it Right:} $L_z$-values of the particles at time 
$2.2$~Gyr within the radial ranges $R=(2.5\pm 0.5)$~kpc (bin~1, blue line), 
   $R=(3.0\pm 0.5)$~kpc (bin~2, red line) and $R=(8.0\pm 0.5)$~kpc
     (bin~3, black line).}
   \label{Fig:kappa1}
\end{figure*}

The methods described in the previous section allowed us to calculate the
diffusion coefficient $D(R,t)$, which is shown in the contour map in the left panel of
of fig.~\ref{Fig:kappa1}.
The bar's corotation radius is also shown as a
dashed white line. We can see that the diffusion coefficient is not
constant in time nor in radius. If the disk is not too hot, $D$
has the largest values $D\sim 0.1~\rm{kpc}^2\, \rm{Myr}^{-1}$ 
near the corotation    
radius of the bar (which increases in time in our simulation), 
where the density is strongly perturbed
by a $m=2$ pattern created by the bar and the transient spiral arms, 
and in the external regions $R > 8$~kpc,
where the density is modulated by a $m=1$ pattern. The stars respond
collectively to these modulations and the process of migration
corresponds to a diffusion in an axisymmetric system. 

From $D$ we can estimate the radial dispersion $\sigma$ in a
rotation period (which is of the order of the diffusion
time-scale), $\sigma = \sqrt{2\, D\, T_{rot}}$. 
If the disk is sufficiently cold, the radial dispersion is high near
the corotation region of the bar, where it can recurrently assume
values of the order of $\sigma \sim 6$~kpc. 
At an intermediate radius, such as $R =
6$~kpc, the radial dispersion is of the order of $\sigma \sim 3$~kpc
and it increases in the external less dense regions where the 
pattern $m=1$ dominates. 
In this external region, the radial migration is very 
high and the number of stars which stay
in a local volume of radius $d\sim 100$~pc around, for example, $R=8$~kpc is very low for times
less than the diffusion time,
since $d  \ll \sigma \sim 5$~kpc. 

\begin{figure*}
   \centering
   \includegraphics[width=5.2cm]{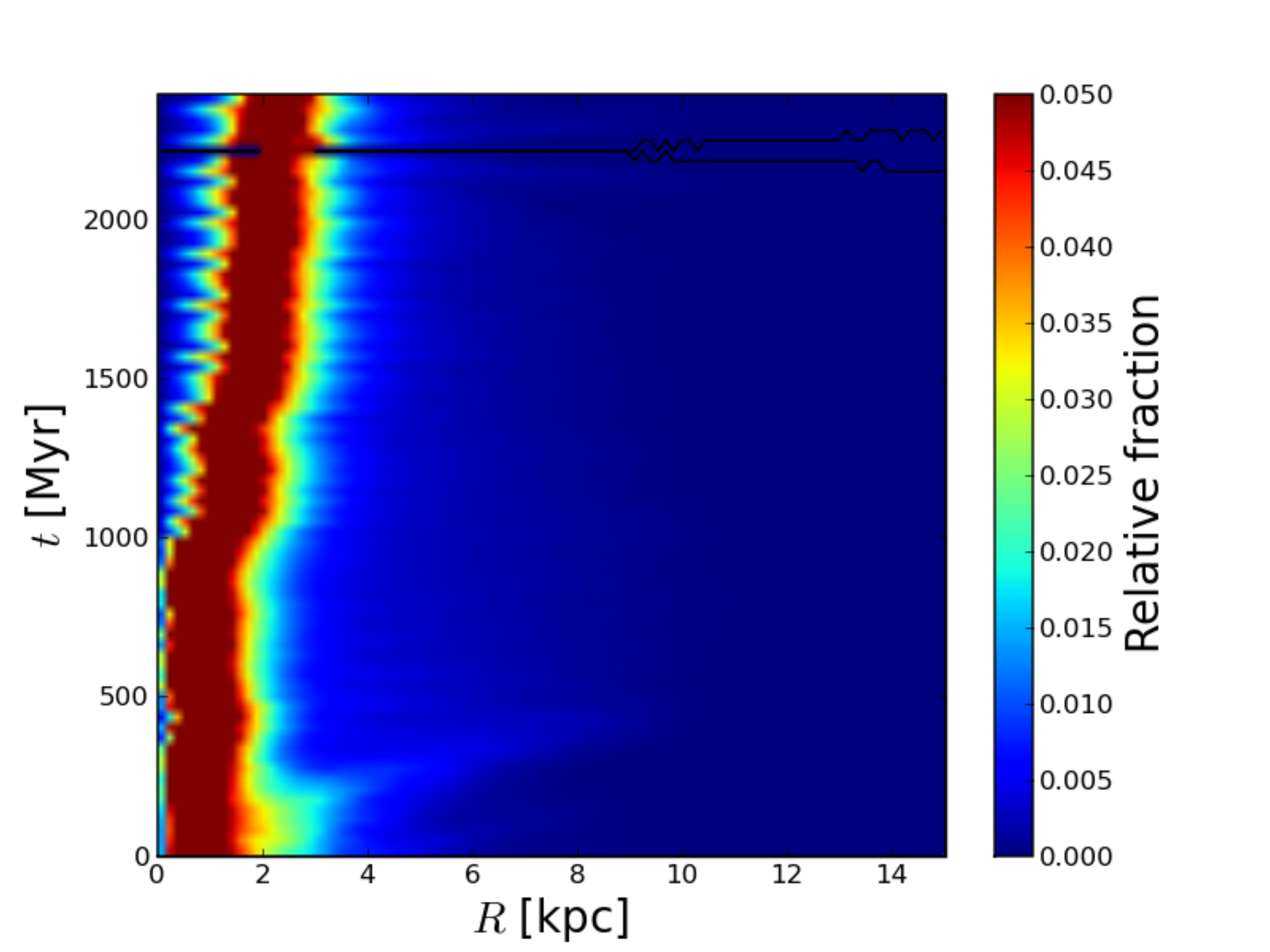} 
   \includegraphics[width=5.2cm]{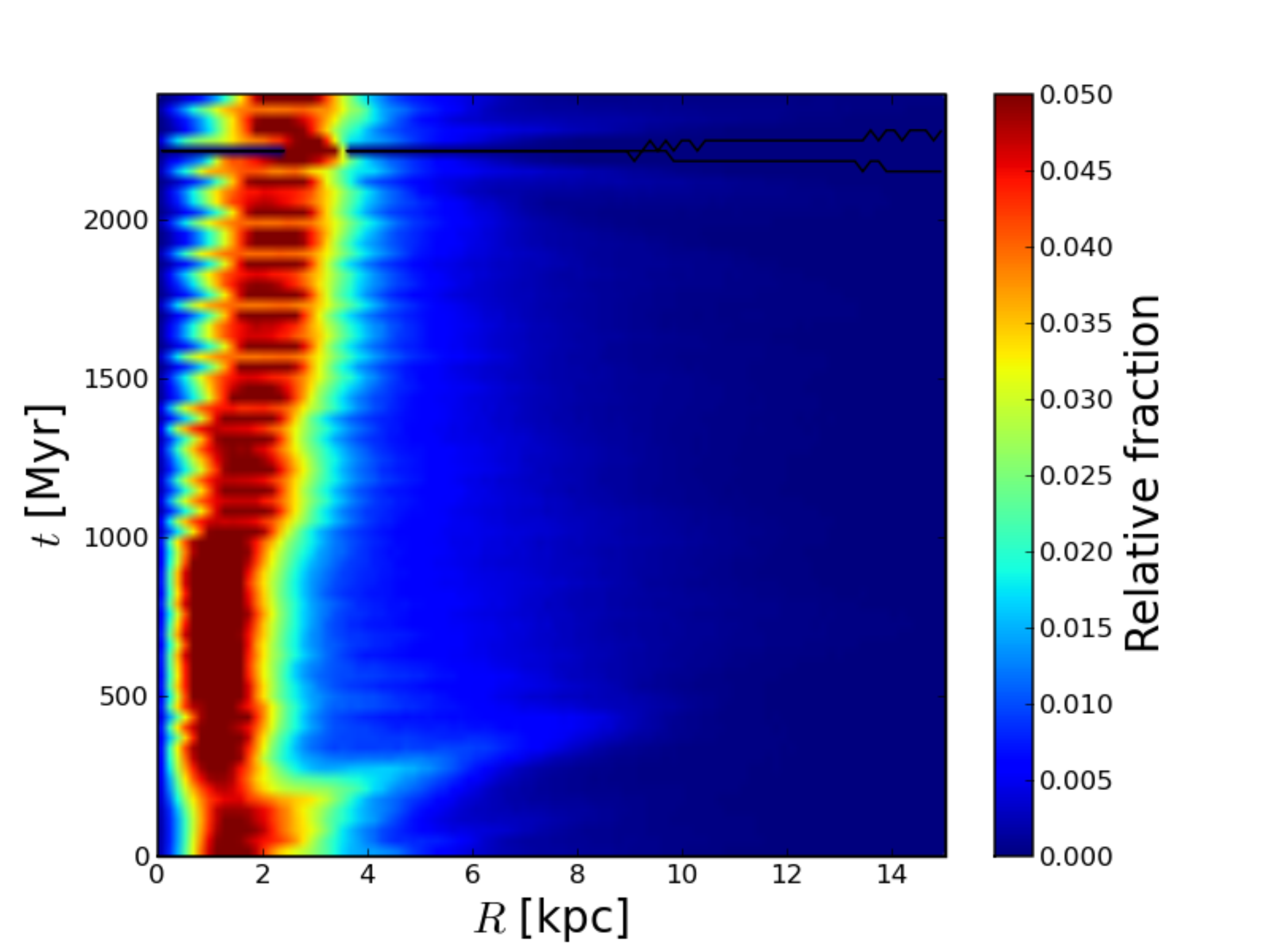} 
   \includegraphics[width=5.2cm]{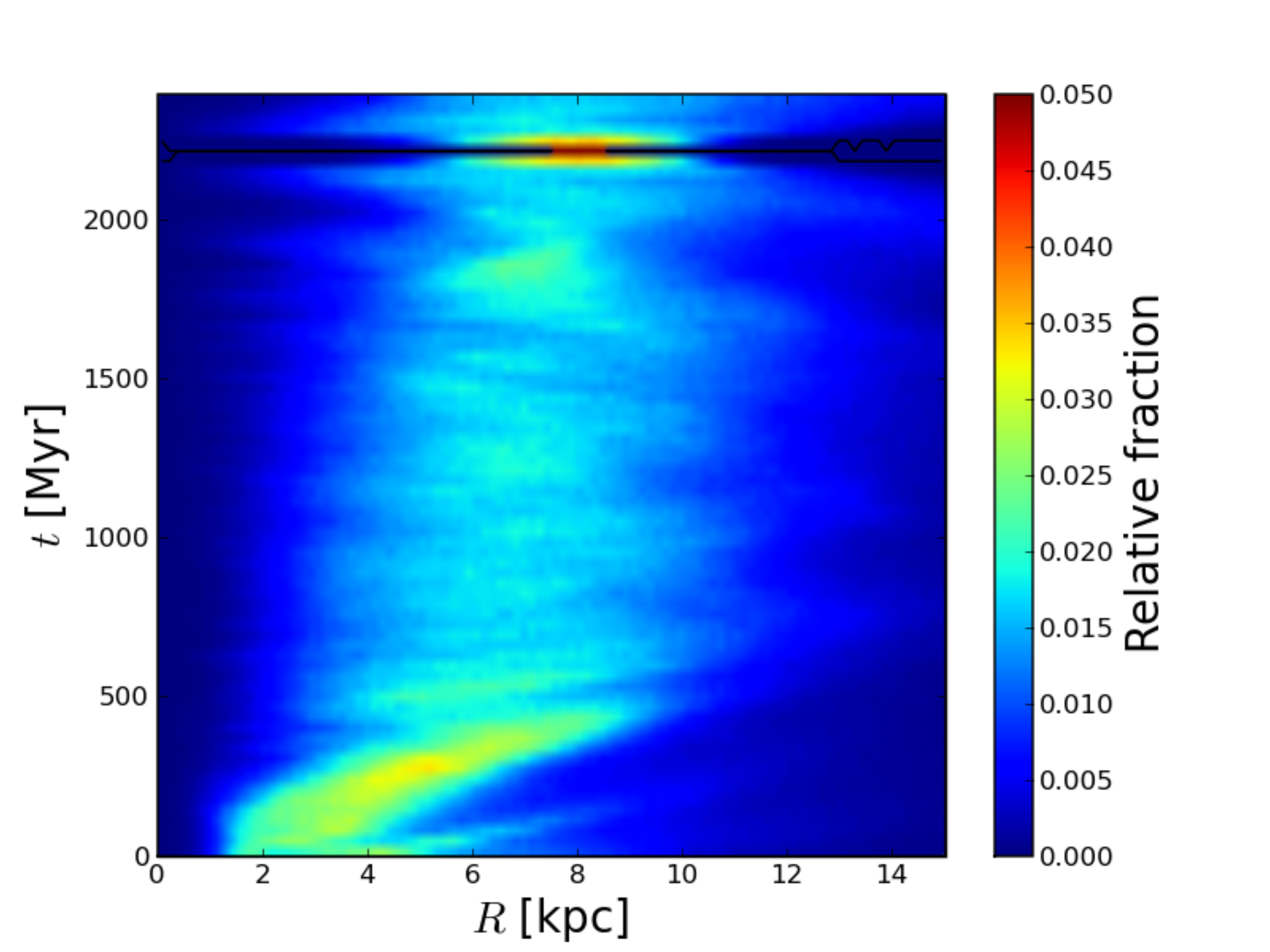} 
   \caption{\small Radial migration of particles which at time $t =
   2.2$~Gyr are in $R=(2.5\pm 0.5)$~kpc ({\it left panel}), 
   $R=(3.0\pm 0.5)$~kpc ({\it middle panel}) and $R=(8.0\pm 0.5)$~kpc 
   ({\it right panel}).}
    \label{Fig:R}
\end{figure*}

\begin{figure*}
   \centering
   \includegraphics[width=5.2cm]{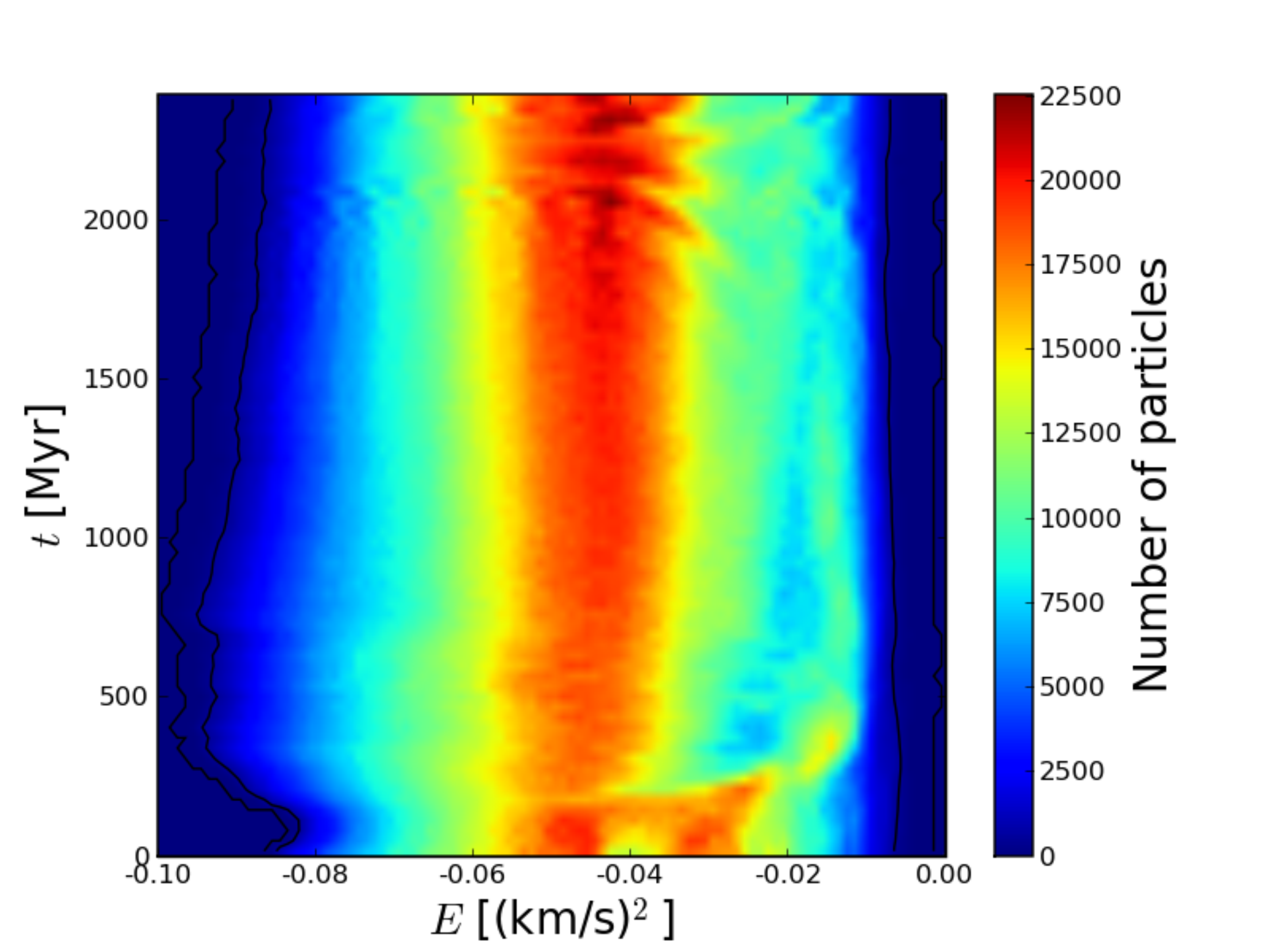} 
   \includegraphics[width=5.2cm]{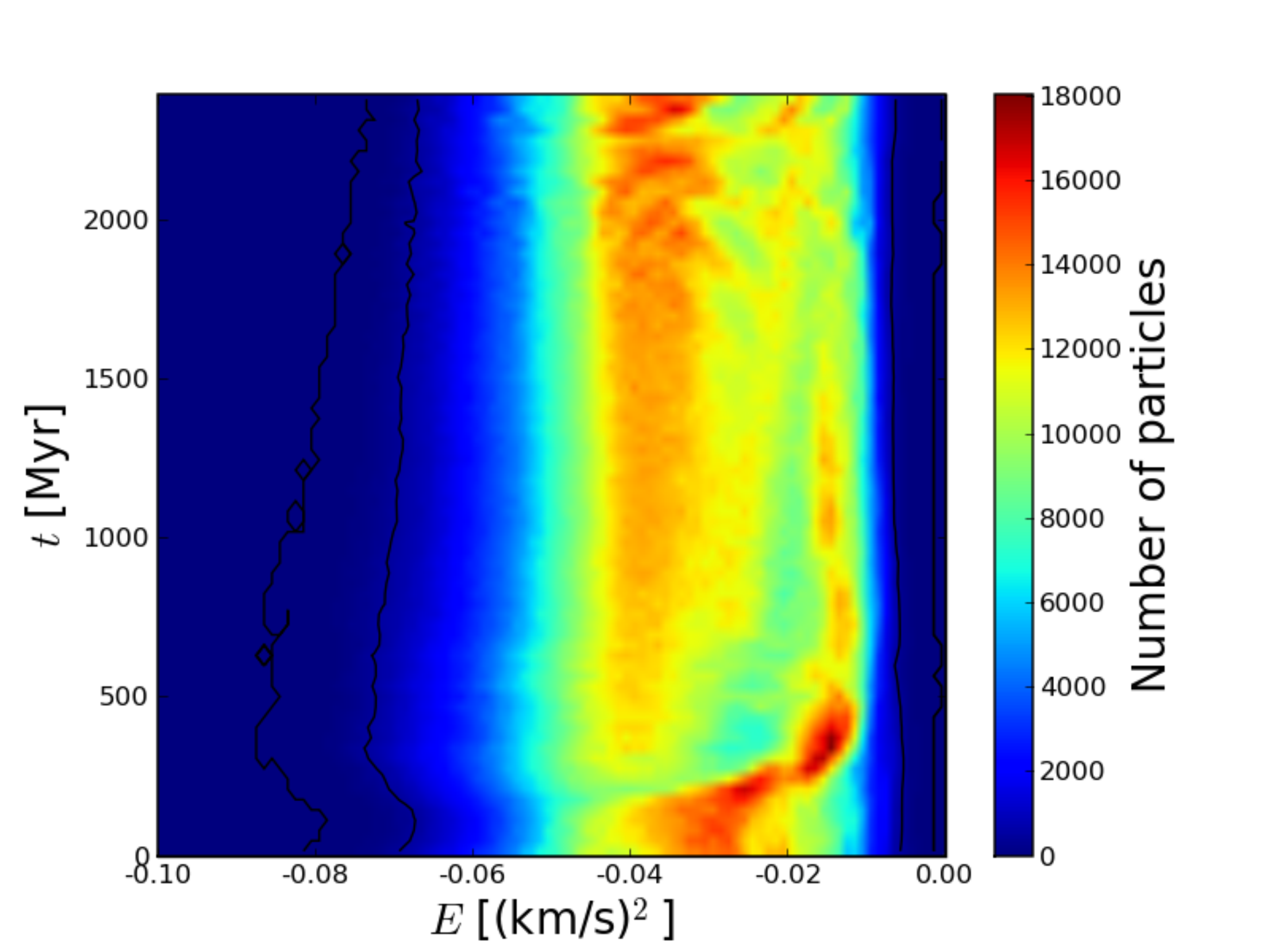} 
   \includegraphics[width=5.2cm]{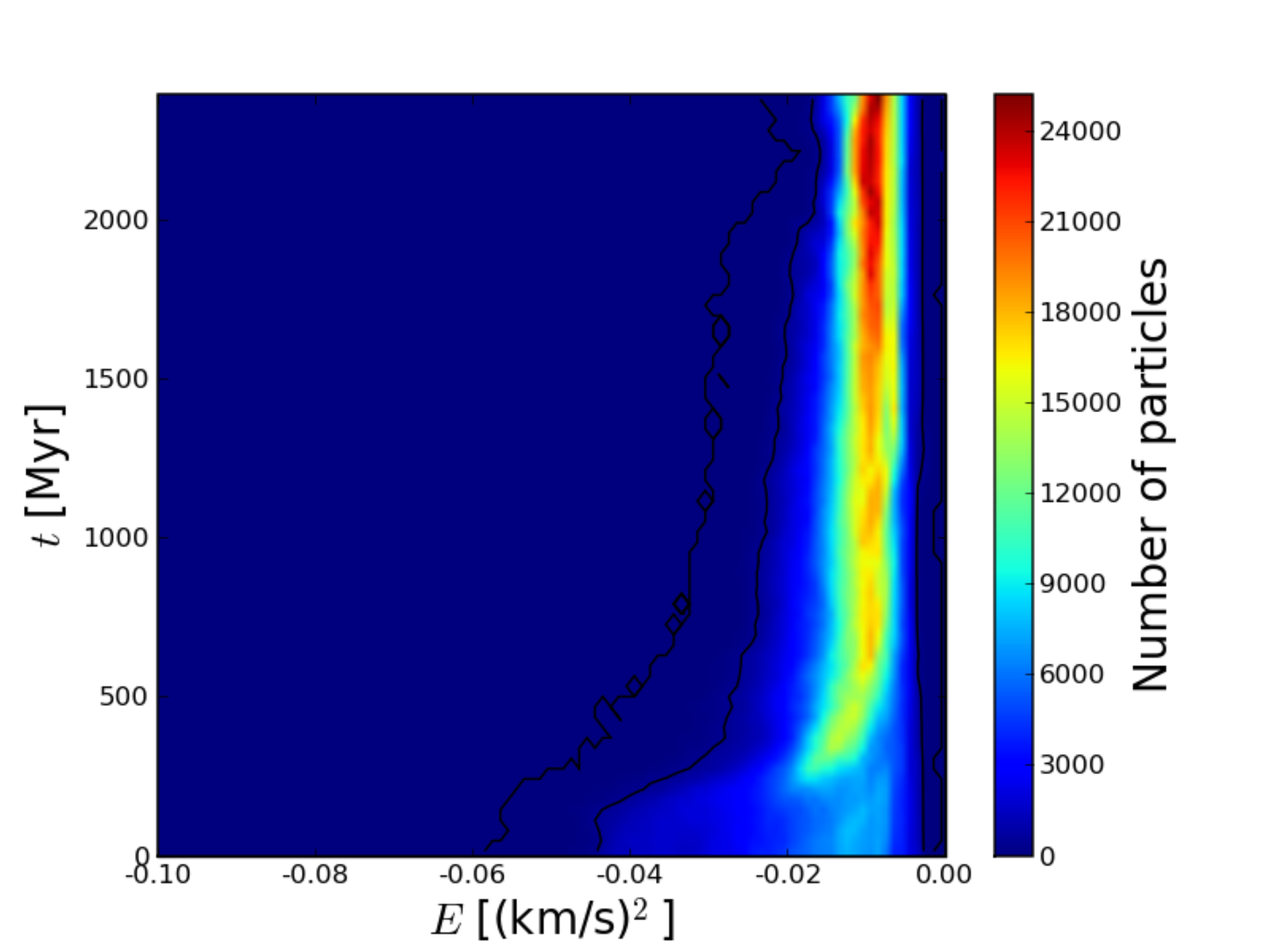} 
   \caption{\small Energies of particles which at time $t =
   2.2$~Gyr are in $R=(2.5\pm 0.5)$~kpc ({\it left panel}),
   $R=(3.0\pm 0.5)$~kpc ({\it middle panel}) and $R=(8.0\pm 0.5)$~kpc 
   ({\it right panel}).}
    \label{Fig:E}
\end{figure*}

Three different bins of particles located respectively in 
$R=(2.5\pm 0.5)$~kpc,  $R=(3.0\pm 0.5)$~kpc, $R=(8.0\pm 0.5)$~kpc at 
$t=2.2$~Gyr are followed in time and their evolution history is shown 
in fig.~\ref{Fig:R}.  In the first bin, particles are inside the bar
and they remain there all along the evolution. The presence of a $m =1$ mode in the central bar region can be observed as periodical modulations in the density. These particles have
values of $L_z \sim 0$ (see the right panel of fig.~\ref{Fig:kappa1}, blue line labeled as `bin 1')
and large negative energies (see the left panel of fig.~\ref{Fig:E}). Particles in the second
bin centered in $R=(3.0\pm 0.5)$~kpc at $t = 2.2$~Gyr belong to two different types of
orbits: one family can migrate only inside, the other can go outside 
the bar, in the disk. Large values of the diffusion coefficient $D$ 
near the corotation region (see the left panel of fig.~\ref{Fig:kappa1}) are related to 
this superposition of two families of bar orbits.  The
corresponding values of $L_z$ and $E$ are shown, respectively, in the red line 
in fig.~\ref{Fig:kappa1} (left panel, line labeled as `bin 2') and in the middle panel of fig.~\ref{Fig:E}, where it can be 
easily seen that the two families have two different ranges of
energies: large negative energies correspond to bar particles and small negative energies 
correspond to disk particles. 
In the third bin, particles belong to the disk component: they can span all the 
region outside corotation, $R_c < R < 12$~kpc, and they come mainly 
from the corotation region at the time when the bar strength had a 
maximum  (i.e. at  $t\sim 350$~Myr, cf. fig.~\ref{RotCurve}d).  These disk particles have small negative energies (see the right panel of fig.~\ref{Fig:E}) and large values of $L_z$ (see the black line labeled as `bin 3' in the right panel of fig.~\ref{Fig:kappa1}).

\section{Conclusions}

Modeling the migration of stars in marginally stable disks as a diffusion process in the radial direction is a powerful tool which allows us to estimate quantitatively two crucial parameters, the diffusion coefficient and the diffusion time-scale. It is important to note that such diffusion model is only valid  
for describing the system for times smaller than the diffusion time-scale, which is of the order of the rotation period, and for times larger than the chaotic time-scale, after which orbits are no more time-reversible. 
 
The calculations of both the diffusion coefficient and the diffusion
time-scale give us a quantitative measure of the migration process in the
disk. Another advantage of studying the diffusion of stars in real
space, rather than in velocity space, is that it can be easily
related to the evolution of chemical elements, thus representing an 
interesting tool to be implemented in chemical evolution codes. 

We have found that the diffusion time-scale is of the order of one
rotation period and that the diffusion coefficient $D$ depends on the
evolution history of the disk and on the radial position. Larger
values $D$ are found near the corotation region, which evolves in time,
and in the external region, where asymmetric patterns develop.  
Marginally stable disks, with $Q_T \sim 1$, have two different
families of bar orbits with different values of angular momentum $L_z$ 
and energy $E$, which determine a large diffusion in the corotation region. 

The application of the method described here to models which include the halo component and have disks with different initial Toomre-parameter $Q_T$ will be presented in a forthcoming paper. 

\smallskip
\noindent
{\bf \small Acknowledgements} 
{\small The simulations were run on the REGOR cluster at Geneva Observatory. This work has been 
supported by the Swiss National Science Foundation.}

\end{document}